\RequirePackage{fix-cm}
\documentclass[twocolumn,epjc3]{svjour3}
\smartqed  

\RequirePackage{graphicx}
\RequirePackage{mathptmx}      
\RequirePackage{flushend}
\RequirePackage[numbers,sort&compress]{natbib}
\RequirePackage[colorlinks,citecolor=blue,urlcolor=blue,linkcolor=blue]{hyperref}

\RequirePackage{latexsym}
\RequirePackage[numbers,sort&compress]{natbib}
\RequirePackage[colorlinks,citecolor=blue,urlcolor=blue,linkcolor=blue]{hyperref}
\begin{document}

\title{Spectroscopy of a  Reissner-Nordstr\"{o}m black hole via an action variable
}


\author{Xiao-Xiong Zeng \thanksref{e1, addr1,addr2}
         \and
Wen-Biao Liu(corresponding author)
        \thanksref{e2, addr1}}

\thankstext{e1}{e-mail: xxzeng@mail.bnu.edu.cn}
\thankstext{e2}{e-mail: wbliu@bnu.edu.cn}
\institute{Department of Physics, Institute of Theoretical Physics,
Beijing Normal University, Beijing 100875, China \label{addr1}
          \and
Department of Physics and Engineering Technology, Sichuan University
of Arts and Science, Dazhou 635000, Sichuan, China \label{addr2} }

\date{Received: date / Accepted: date}

\maketitle

\begin{abstract}
With the help of the Bohr--Sommerfeld quantization rule,  the area
spectrum of a charged, spherically symmetric
 spacetime is obtained by studying an adiabatic invariant action  variable. The period of the Einstein-Maxwell
 system, which is  related to the surface gravity of a given spacetime, is determined by the Kruskal--like coordinates.
  It is shown that the  area spectrum of the Reissner-Nordstr\"{o}m black hole is  evenly spaced and the spacing is the
  same as that of a Schwarzschild black hole, which indicates that the area spectrum of a black hole is independent
  of its parameters. In contrast to the quasi-normal mode analysis, we do not  impose the small charge limit as the general
area gap $8\pi $ is obtained.

\end{abstract}

\section{Introduction}

Investigation the properties of black holes has attracted attention
of more and more astronomers and physicists in the past several
years. On one hand, people want to uncover the mystery of   black
holes by detecting  their  fingerprints. On the other hand, black
holes may be a base to test any scheme for a quantum theory of
gravity. A crucial development in black hole physics is the
discovery of Hawking \cite{Hawking1974} that a black hole was not
totally black  but can emit radiation with Hawking temperature
$T_{BH}=\hbar\kappa_+/2 \pi$, where $\kappa_+ $ is the surface
gravity at the event horizon. Based on the  Hawking temperature, one
can easily obtain  the well known Bekenstein--Hawking entropy
$S_{BH}=A/4l_p^2$, where $A$ is the horizon area and $ l_{p}=(G
\hbar / c^3)^{1/2}$ is the Planck length.

The idea of quantization of a black hole  was proposed first  by
 Bekenstein \cite{Bekenstein1974,Bekensteingr-qc/9710076,Bekensteingr-qc/9808028}, and was
 motivated by the analog that a black hole plays the role in gravitation as an atom plays in the quantum mechanics.
He observed that the horizon area of a non-extremal black hole is
classically adiabatic invariant. According to the Ehrenfest
principle, any classical adiabatic invariant corresponds to a
quantum entity with discrete spectrum; Bekenstein hence conjectured
that the horizon area of a non-extremal quantum black hole has a
discrete eigenvalue spectrum. Based on Christodoulou's point
particle model \cite{Christodoulou1970}, Bekenstein found that the
smallest possible increase in horizon area of a non-extremal black
hole is $ \Delta A=8\pi l_{p}^2$, as the Heisenberg quantum
uncertainty principle was considered \cite{Bekensteingr-qc/9710076}.
 Following the pioneering
work of Bekenstein \cite{Bekensteingr-qc/9710076}, a lot of
independent calculations have been done and the uniformly spaced
area spectrum was  reproduced
\cite{Louko1996,Makela1997,Dolgov1997,Peleg1995,Kastrup1996}. Later
on, Hod \cite{Hod1998} elucidated that the analysis of Bekenstein
was only analogous to the well-known semiclassical determination of
a lower bound on the ground state energy of the hydrogen atom, and
that instead one should consider a wave analysis of black hole
perturbations. In terms of Bohr's correspondence principle, Hod
found that the area spectrum of a black hole can
 be determined by the analysis of the asymptotic behavior (i.e., n$\longrightarrow \infty$) of the highly damped quasi-normal
  mode frequencies. Using the real part of the quasi-normal mode frequency of the  Schwarzschild black hole,  the area
  spectrum was found to be  $\Delta A=4 \lambda l_{p}^2 $, in which $\lambda =\ln 3$. In 2002, Kunstatter \cite{Kunstatter2003}
   further confirmed this result,  combining the proposal of Bekenstein that the black hole horizon area is adiabatic invariant and the  proposal
    of Hod that the quasi-normal mode frequency is responsible for the area spectrum. According to  Kunstatter's viewpoint, the ratio
     $M/\omega$ is invariant, as the quasi-normal mode frequency $\omega$ and  black hole mass $M$ are treated as the classical vibrational
      frequency and system energy in large $n$ limit.
On the basis of
 statistical interpretation for black hole entropy, Bekenstein and Mukhanov \cite {Bekenstein19952} found that the
parameter $\lambda$  should be equal to $\ln k$,  so the results of
Hod and Kunstatter  are  believed to be  more reasonable physically.
  In 2007, Maggiore \cite{Maggiore2008} gave a new interpretation of the black hole quasi-normal mode frequencies. He found that the proper
   frequency of the equivalent harmonic oscillator,
which is interpreted as the quasi-normal mode frequencies
$\omega=\omega_R+i \omega_I$, should consider contributions of both
the real part
 $\omega_R$ and imaginary part $\omega_I$ in  high damping limit. More importantly, Maggiore found that the most interesting case is that
 for the highly excited quasi-normal modes, where the imaginary part  rather than the real part is dominant. In this case,  the area spectrum
  of a Schwarzschild black hole is found to be $ \Delta A=8\pi l_{p}^2$, which is consistent with the semiclassical result of Bekenstein.  Based on
the ideas of Hod, Kunstatter, and Maggiore, the area spectrum and
entropy spectrum of many black hole spactimes have been investigated
\cite{all}. It was shown that the  spacings for the entropy spectrum
and area spectrum are equidistant in Einstein gravity. In modified
gravity theory, it was found that the entropy spectrum is
equidistant, but the area spectrum is not equidistant
\cite{Kothawala2008}. As  higher order quantum corrections to the
semiclassical action are considered \cite{jiang2010}, the entropy
spectrum and area spectrum obey the same rules as in modified
gravity theory.

Very recently, Majhi and Vagenas \cite{Majhi2011} proposed another
scheme to study black hole spectroscopy. In their work, they did not
resort to the quasi-normal mode frequencies of a black hole to find
the area spectrum. Instead, they  used an adiabatic invariant $\int
{p_i dq_i}$, which obeys the  Bohr-Sommerfeld quantization rule;
 here $p_i$ is the conjugate momentum of the coordinate $q_i$. Their idea was motivated mostly by the initial inference of Bekenstein
 that the horizon area of a black hole is adiabatic invariant.
   As the period of the gravity system is given, which was shown to be related to the surface gravity at the event horizon of the black hole,
  they obtained an equally spaced entropy spectrum of a static, spherical symmetry black hole with
its quantum to be equal to the one given by Bekenstein.

In this paper, we intend to investigate the spectroscopy of a
Reissner-Nordstr\"{o}m (RN) black hole. For the area spectrum of a
charged black hole,
 there have been many investigations. In terms of the reduced phase-space quantization,  Barvinsky et al \cite{Barvinsky2001} found that
 the horizon area
should be $A_{ n,p} = 4\pi (2n + p + 1) l_p^2$,  where $n,
p=0,1,2,\cdots$ and the quantum number $p$ corresponds to
$Q=\pm\sqrt{\hbar p}$. Making use of Bohr's correspondence
principle,  Hod showed that the  area spectrum was  $ \Delta A=4 \ln
2 l_p^2$ and $ \Delta A=4 \ln 3 l_p^2$ respectively, for the event
horizon area and the total areas of the inner horizon and outer
horizon \cite{Hod2006,Hod2007}. Recently, Banerjee
 et al \cite{Banerjee279} got the value  $ \Delta A=4  l_p^2$ from the viewpoint of the tunneling paradigm. Based on
  the quasi-normal modes analysis, Wei et al \cite{Wei2010} and  Lopez-Ortega \cite{Lopez-Ortega}, respectively, obtained
  the formalism $ \Delta A=8\pi l_{p}^2$ for the  charged Garfinkle-Horowitz-Strominger black
hole and  D-dimensional RN black hole. Note that in Refs.
\cite{Wei2010,Lopez-Ortega}, the  small charge limit was imposed. It
is obvious that there is no unanimous consensus  on the spacing of
area spectrum of a  charged black hole.
 According to the viewpoint of Bekenstein \cite{Bekensteingr-qc/9710076}, the spacing is related to the  total number of quantum states
 of the horizon; therefore an -n-depth understanding the spacing of area spectrum of a black hole  may shed light on the statistics origin
 of black hole entropy or information.

Motivated by Ref. \cite{Majhi2011}, we will investigate area
spectrum of an RN black hole using an adiabatic invariant action
variable.  In analytical mechanics, it is known that the action $I$,
action variable  $I_v $, and
Hamiltonian ${H}$ of any single periodic system  satisfy the relation $%
I= I_v- \int{H} dt$. Thus as the action and Hamiltonian are given,
the quantized action variable can be  obtained with the help of the
Bohr-Sommerfeld quantization rule. We will prove that the quantized
action variable is nothing but the entropy of the black hole, and
thus the entropy and further the horizon area of the black hole can
be quantized. In the literature,  it is indicated that the quantum
effects of interest take place at  the event horizon and that the
physics near the horizon can be described effectively by a
two-dimensional spacetime  \cite{Robinson2005}. We hence restrict
our investigation to the effective two-dimensional background.

 The remainder of this paper is arranged as follows. In Sec.2, we will find the period of the Einstein-Maxwell system. In Sec.3,
  on the basis of the Bohr-Sommerfeld quantization rule, we will discuss the spacings of the entropy spectrum and area spectrum by
  studying the adiabatic invariant action variable.  Sec.IV presents our  conclusion.

\section{Reissner-Nordstrom black hole in Kruskal-like coordinates}
 The line element of an RN black hole is%
\begin{equation}
 \label{1} ds^{2}=-f(r)dt^{2}+f^{-1}(r)dr^{2}+r^{2}d\theta ^{2}+r^{2}\sin^{2}\theta d\varphi
^{2},
\end{equation}
where%
\begin{equation}
 \label{2}f(r)=1-\frac{2M}{r}+\frac{Q^{2}}{r^2}=\frac{(r-r_+)(r-r_-)}{r^2},  \label{2}
\end{equation}%
in which $r_{\pm}=M\pm\sqrt{M^{2}-Q^{2}}$ are the outer horizon and inner horizon and $M,Q$ are the mass and charge of the black hole. The
electromagnetic potential is%
\begin{equation}
 \label{3}A_{\mu }=(\frac{Q}{r},0,0,0).  \label{3}
\end{equation}%
 More and more literature publications have indicated that the interesting quantum phenomena, such as Hawking radiation,
take place at the event horizon of  a black hole, and  near the
horizon the higher dimensional
  background can effectively be reduced to a two dimensional spacetime  \cite{Robinson2005}. In this paper, we are going to
  investigate the change in horizon area of the
  RN black hole while a particle runs out from the event horizon,
   so we restrict our study to the  effective two dimensional spacetime.

According to the dimensional reduction technique, the
two-dimensional spacetime of an RN black hole can be written as
\cite{Iso20063}
\begin{equation}
 \label{4}ds^{2}=-f(r)dt^{2}+f^{-1}(r)dr^{2}.  \label{4}
\end{equation}%
In this spacetime, it is well known that  the event horizon is a
coordinate singularity. To get the period of the Einstein-Maxwell
system, we should introduce the Kruskal-like  coordinate. Without
losing the integrity of this paper, we will give some key steps to
construct the Kruskal-like  coordinates.

 As the first step of the coordinate transformations, we use the tortoise coordinate defined
as
\begin{equation}
dr^{\ast }=\frac{1}{f(r)}dr.  \label{5}
\end{equation}
Integrating it over $r$ from $0$ to $r$, we obtain
\begin{equation}
r^{\ast }=r+\frac{1}{2\kappa_+}\ln \frac{r-r_+}{r_+}+\frac{1}{2\kappa_-}\ln \frac{r-r_-}{r_-}, \label{6}
\end{equation}
in which $\kappa_\pm =\frac{r_{+}-r_-}{2 r_{\pm}^{2}}$ is the
surface gravity on the outer (inner) horizon. Note that
 we have considered only the range $r>r_+ $ here.
Using the  null coordinates $u=t-r^{\ast }$, $v=t+r^{\ast },$ we can
construct the coordinates $U=-e^{-{\kappa_+ u}}$, $%
V=e^{\kappa_+ v}$. In this case, Eq. (\ref{4}) can be rewritten as
\begin{equation}
ds^{2}=-\kappa_+^{-2} e^{\kappa_+(\mu-\nu)}f(r)dUdV.  \label{7}
\end{equation}%
Defining
\begin{equation}
T=\frac{1}{2}(V+U)=e^{\kappa_+ r}(\frac{r-r_+}{r_+})^{\frac{1}{2}}
(\frac{r-r_-}{r_-})^{\frac{\kappa _+}{2\kappa _-}} \sinh\kappa_+ t,
\label{8}
\end{equation}
\begin{equation}
R=\frac{1}{2}(V-U)=e^{\kappa_+ r}(\frac{r-r_+}{r_+})^{\frac{1}{2}}
(\frac{r-r_-}{r_-})^{\frac{\kappa _+}{2\kappa _-}} \cosh\kappa_+ t,
\label{9}
\end{equation}%
Eq. (\ref{7}) can be further expressed as
\begin{equation}
ds^{2}=\kappa_+^{-2} e^{-2 \kappa_+r}(\frac{r-r_-}{r_+})(\frac{r_-}{r-r_-})^{\frac{\kappa_+}{\kappa_-}}(-dT^2+dR^2),  \label{10}
\end{equation}%
in which $T,R$ are the Kruskal coordinates. Extending the time
coordinate to its imaginary axis, namely $t=-i\tau$, Eq. (\ref{8})
and Eq. (\ref{9}) can be rewritten as
\begin{equation}
iT=\frac{1}{2}(V+U)=e^{\kappa_+ r}(\frac{r-r_+}{r_+})^{\frac{1}{2}} (\frac{r-r_-}{r_-})^{\frac{\kappa _+}{2\kappa _-}} \sin\kappa_+ \tau,  \label{11}
\end{equation}
\begin{equation}
R=\frac{1}{2}(V-U)=e^{\kappa_+ r}(\frac{r-r_+}{r_+})^{\frac{1}{2}} (\frac{r-r_-}{r_-})^{\frac{\kappa _+}{2\kappa _-}} \cos\kappa_+ \tau.  \label{12}
\end{equation}%
Obviously, both $T,R$ are periodic functions with respect to the
Euclidean time $\tau$ with period ${2\pi}/\kappa_+$. So the period
for any continuous function on variable  $T,R$ is ${2\pi}/\kappa_+$
too. This period is very useful for  studying Hawking temperature
from the viewpoint of the temperature Green function
\cite{Gibbons1978}. Next, we will
 use this period to study the area spectrum of the RN black hole  via an adiabatic invariant action variable.

\section{Entropy spectrum and area spectrum}
The adiabatic invariant is a notion in analytical mechanics. A
physical system governed by a Hamiltonian $H (q, p, \lambda(t))$ is
regarded as undergoing an adiabatic change if $\lambda(t)$  varies
on a timescale longer than the longest timescale $T$  of the
internal motions, namely $T\frac{d \lambda}{dt}\ll \lambda$.
 According to the view of Ehrenfest, all Jacobi action integrals of the form $I_v=\oint p dq $ for a quasi-periodic system
are  adiabatic invariant action variable. In this section, we will
employ the adiabatic invariant action variable $I_v$, which obeys
the Bohr-Sommerfeld quantization rule $I_v=nh$,  to investigate area
spectrum of an RN black hole.

It is well known that in classical mechanics, the  action $I$,
action variable
 $I_v $,  and  the  Hamiltonian  ${H}$ of  any
single periodic system satisfy the relation
\begin{equation}
I= I_v- \int{H}  dt.    \label{14}
\end{equation}%
Obviously, to obtain the action variable, one should find the
Hamiltonian of the system and the action of  the moving particle.
For any classical Einstein gravity system, the Hamiltonian
(including its Gibbons-Hawking surface term) is always the
 Arnowitt-Deser-Misner (ADM) mass of the system. Based on
Eq. (\ref{14}), it was found that there is a ground state for the
Schwarzschild black hole, which may be a candidate of dark matter
\cite{Liu2004}.

For the charged background spacetime, the effect of the
electromagnetic field on action variable must be considered.
Firstly, we concentrate on the Hamiltonian.
 To investigate the first law of black hole mechanics, Wald \cite{Wald1993,Sudarsky1992} once obtained the Hamiltonian for the
 Einstein-Maxwell system. It was found that
 the ADM Hamiltonian for Einstein-Maxwell theory has the
form
\begin{equation}
H=\int_{\Sigma}(N^{\mu}C_{\mu}+N^{\mu}A_{\mu}C), \label{16}
\end{equation}%
where $\Sigma$ is  a three-dimensional manifold, $N^{\mu}$,  $A_{0}$
are  non-dynamical variables, and
\begin{equation}
0=C=\frac{1}{4\pi}\sqrt{h}D_aE^a,  \label{17}
\end{equation}
 \begin{equation}
0=C_0=\frac{1}{16\pi}\sqrt{h}[-R+2E_aE^a+F_{ab}F^{ab}+\frac{1}{h}(\pi_{ab}\pi^{ab}-\frac{1}{2}\pi^2)], \label{18}
\end{equation}
\begin{equation}
0=C_a=-\frac{1}{8\pi}\sqrt{h}[D_b(\pi^b_{a}/\sqrt{h})-2F_{ab}E^b], \label{19}
\end{equation}
in which $E^a$ is the electric field, $\pi^{ab}$ is the momentum
canonically conjugate to $h_{ab}$, and $F_{ab}=2D_{[a}A_{b]}$ is the
field strength.
 In the case where $N^{\mu}$ asymptotically approaches
a time transition and considering  contributions from boundary terms
at infinity, the  Hamiltonian can be written as
\begin{equation}
H=H_{\Sigma}+\frac{1}{16\pi}\oint_{\infty}dS^a[\partial^bh_{ab}-\partial_a h^{b}_a]+\frac{1}{4\pi}\oint_{\infty}dS^aA_0E^a.  \label{20}
\end{equation}
Taking into account the classical Hamiltonian  constraints,
$H_{\Sigma}$ will vanish, and hence the  Hamiltonian  should be
\begin{eqnarray}
H&=&\frac{1}{16\pi}\oint_{\infty}dS^a[\partial^bh_{ab}-\partial_a h^{b}_a]+\frac{1}{4\pi}\oint_{\infty}dS^aA_0E^a \nonumber \\
&=&M+\phi^{\infty} Q,  \label{21}
\end{eqnarray}
in which $\phi=Q/r$ is the electrostatic potential and ``$\infty$"
stands for the value at infinity.  It should be pointed out that the
gauge potential in Eq. (\ref{3}) is divergent in the usual RN
coordinates. So a transformation must be performed so that the
vector potential will smooth through the horizon. In Ref.
\cite{Gao044016}, the transformation
$A_a^{\prime}=A_a+\frac{Q}{r_+}(dt)_a$ has been used. Here we also
adopt this skill. In this case, the
 Hamiltonian changes to
\begin{equation}
H=M-\phi^{+} Q,  \label{22}
\end{equation}
where $\phi^{+}=Q/r_+$ is the electrostatic potential at the event
horizon.

 For the action,  many works have been published
\cite{all2}. It has been shown that
 the  general coordinate $A_{\mu}=(A_{t},0)$ is an ignorable coordinate and the
corresponding  freedom should be eliminated completely. In this
case, the action $I$ can be written as
\begin{equation}
I= \int p_r dr-\int p_{A_t} dA_t,   \label{23}
\end{equation}%
where $p_r$, $p_{A_t}$ are the  canonical momenta conjugate to $r$
and $A_t$, respectively. Therefore, for the two-dimensional
Euclidean metric
\begin{equation}
ds^{2}=f(r)d{\tau}^{2}+f^{-1}(r)dr^{2},  \label{24}
\end{equation}
the invariant action variable can be expressed as
\begin{eqnarray}
\int{H}d\tau&+&\int p_r dr-\int p_{A_{\tau}} dA_{\tau} \nonumber \\
&=&\int{H}d\tau+\int \int _0^{p_r}dp^{\prime}_r dr-\int
\int _0^{p_{A_{\tau}}}dp^{\prime}_{A_{\tau}} dA_{\tau},    \label{25}
\end{eqnarray}
in which $\tau=-it$ is the  Euclidean time. Considering  the
following Hamilton's equations:
\begin{equation}
\dot{r}=\frac{dH}{dp^{\prime}_r}\mid_{(r;A_{\tau},p_{A_{\tau}})},  \label{26}
\end{equation}
\begin{equation}
\dot{A_{\tau}}=\frac{dH}{dp^{\prime}_{A_{\tau}}}\mid_{(A_{\tau};r,p_{A_{\tau}})},  \label{27}
\end{equation}
 the invariant action variable can be rewritten as
\begin{eqnarray}
\oint p dq &=&\int{H}d\tau+\int \int _0^{H_{(r;A_{\tau},p_{A_{\tau}})}}\frac{dH_{(r;A_{\tau},p_{A_{\tau}})}^{\prime}}{\dot{r}} dr  \nonumber \\
&-&\int
\int _0^{H_{(A_{\tau};r,p_{A_{\tau}})}}\frac{dH_{(A_{\tau};r,p_{A_{\tau}})}^{\prime}}{\dot{r}} dr.   \label{28}
\end{eqnarray}
 To obtain  $\dot{r}$, one should consider the  radial,  null
geodesics while a particle runs out.  This method has been used  to
study the tunneling effect extensively \cite{all2}.
 Solving Eq. (\ref{24}), we get
\begin{equation}
\dot{r}\equiv\frac{dr}{d\tau}=\pm if(r),   \label{29}
\end{equation}%
where  $\pm$ denotes the outgoing (incoming) radial null paths. Thereafter,
 we only focus on the outgoing paths, since these are the ones related to the
quantum mechanically nontrivial features under consideration. In
this case, Eq. (\ref{28})  change to
\begin{equation}
\oint p dq=\int{H}d\tau+\int \int _0^{M}dM^{\prime} d\tau-\int\int _0^{Q}\phi dQ^{\prime}d\tau,    \label{30}
\end{equation}%
where we have used
\begin{equation}
dH_{(r;A_{\tau},p_{A_{\tau}})}^{\prime}=dM^{\prime},    \label{31}
\end{equation}%
\begin{equation}
dH_{(A_{\tau};r,p_{A_{\tau}})}^{\prime}=\phi dQ^{\prime}.    \label{32}
\end{equation}%
Incorporating Eq. (\ref{22}), Eq. (\ref{30})  can be further
simplified as
\begin{equation}
\oint p dq= 2\int\int_{(0,0)}^{(M,Q)}(dM^{\prime}-\frac{Q}{r_+}dQ^{\prime})d\tau.    \label{33}
\end{equation}%
 Since we are considering only the outgoing paths, the integration limit for $\tau$ should be
 $[0,\frac{\pi}{\kappa_+}]$\cite{Majhi2011}. Using the relation $T_{BH}=\frac{\hbar \kappa_+}{2\pi}$, we can  obtain
 \begin{equation}
\oint p dq=\hbar
\int_{(0,0)}^{(M,Q)}\frac{1}{T_{BH}}(dM^{\prime}-\frac{Q}{r_+}dQ^{\prime}).
\label{add6}
\end{equation}%
Simplifying Eq. (\ref{add6}), we get
\begin{equation}
\oint p dq= \hbar S_{BH},    \label{34}
\end{equation}
 where the first law of black hole thermodynamics, $dM=T_{BH}dS+\phi_+ dQ$, has been used. Implementing the Bohr-Sommerfeld quantization rule
\begin{equation}
\oint p dq= 2 \pi n\hbar,    \label{35}
\end{equation}
the black hole entropy spectrum can be given directly as
\begin{equation}
S_{BH}= 2 \pi n,   n=1,2,3,\ldots  \label{36}
\end{equation}
One can also get  the spacing for the entropy spectrum
\begin{equation}
\Delta S_{BH}= 2\pi(n+1) - 2\pi{n}=2\pi. \label{37}
\end{equation}
Obviously the entropy of an RN black hole is discrete and the
spacing is equidistant. Recalling that the black hole entropy is
proportional to its horizon area
\begin{equation}
 S_{BH}= \frac{A}{4l_p^2}, \label{38}
\end{equation}
we can also  get the  area spectrum
\begin{equation}
\Delta A= 8 \pi l_p^2. \label{39}
\end{equation}
It is evident that the area spectrum of the RN  black hole is also
equidistant. Our result confirms the initial proposal of Bekenstein
further that the area spectrum is independent of the black hole
parameters and the spacing is $ 8 \pi l_p^2$.

For the spacing of the area spectrum of a black hole,  Medved
 \cite{Medved1005.2838} argued that $8 \pi $ is, by far, the most
qualified candidate  on the basis of the recent tunneling framework
in Ref. \cite{Banerjee2010}. His argument is supported by the
emergent gravity proposed by Verlinde \cite{Verlinde2011}.   To
obtain  Newton's second law of mechanics and Newton's law of
gravitation from thermodynamics, one finds that the unique choice of
the minimal change in the entropy  is $\Delta S=2\pi$, namely, the
area gap should be $8 \pi $. Our result, obviously, agrees with this
argument.

\section{Conclusions}

The area spectrum of  an RN black hole was investigated with the
help of the Bohr-Sommerfeld quantization rule. We obtained an
equally spaced area spectrum by studying the action variable of the
Einstein-Maxwell system. Our result is the same as the one for a
Schwarzschild black hole \cite {Majhi2011}, which confirms the
proposal of Bekenstein that the area spectrum of a black hole is
independent of its parameters. In our calculation, the quasi-normal
mode frequency is not used,  so there is  no confusion as to whether
the real part or imaginary part is responsible for the area
spectrum. This is more convenient and simple. More importantly,
  the small charge limit for a Reissner-Nordstr\"{o}m black hole was not imposed, which is necessary
   from the viewpoint of quasi-normal mode analysis. Also, we should point out the difference of this paper compared with
   the work in Ref. \cite {Majhi2011}, where the area spectrum of black holes was investigated using an adiabatic invariant.
  The relation between the action
variable  and adiabatic invariant has been discussed recently in
Ref.\cite {Kwon27}. It was found that only the action variable can
be quantized as the equally spaced form $I_v=2 \pi n \hbar$, though
the adiabatic invariant can  also be quantized.  For example, if the
action variable $I_v$ can be quantized via Bohr-Sommerfeld
quantization, $I_v^2$  is not equally  spaced even though it is  an
adiabatic invariant. Hence we think the action variable used in this
paper is more reasonable

Our study is also valid for the rotating black hole. For a Kerr
black hole,
 there is an ergosphere between the outer horizon and infinite redshift surface, so matter dropping into
 this region will be dragged inevitably.  To avoid the dragging effect, one should often perform a  dragging coordinate transformation. Recently, it was shown that the ergosphere outside
 a
 Kerr black hole is similar to the electromagnetic field  outside an RN black hole in the reduced
 two-dimensional metric.  The evidence stems from the investigation of  Iso et  al \cite{Iso2006} that
 the partial wave of quantum fields in the four-dimensional rotating black hole background can be interpreted as
a (1+1)-dimensional charged field with a charge proportional to the
azimuthal angular momentum $m$. Thus the area spectrum of a Kerr
black hole can be investigated by the method used for an RN black
hole.

\section*{Acknowledgements}

This research is supported by the National Natural Science
Foundation of China (Grant Nos. 10773002, 10875012, 11175019). It is
also supported by the Fundamental Research Funds for the Central
Universities under Grant No. 105116.




\end{document}